\documentclass[amsmath, amssymb, preprintnumbers, showpacs, showkeys,aps,prl,superscriptaddress,twocolumn]{revtex4-1}

\usepackage{color} 
\usepackage{hyperref}
\usepackage{slashed}
\usepackage{graphicx}
\usepackage{amsmath}
\usepackage{latexsym}
\usepackage{epstopdf}
\usepackage{amsmath}
\usepackage{enumitem}
\usepackage{amssymb,amsmath}
\usepackage{multirow}
\usepackage{mathtools}
\usepackage{bbm}
\usepackage{bm}
\usepackage{amsfonts}
\usepackage{amssymb}
\usepackage{calligra}
\usepackage{calrsfs}
\usepackage{makecell}
\usepackage{hyperref}
\usepackage[normalem]{ulem}

\newcommand{\beq}{\begin{equation}}
\newcommand{\eeq}{\end{equation}}

\newcommand{\be}{\begin{equation}}
\newcommand{\ee}{\end{equation}}

\begin{document}

\title{Superfluid filaments of dipolar bosons in free space}

\author{Fabio Cinti}
\email{cinti@sun.ac.za}
\affiliation{National Institute for Theoretical Physics (NITheP), Stellenbosch 7600, South Africa}
\affiliation{Institute of Theoretical Physics, Stellenbosch University, Stellenbosch 7600, South Africa}
\author{Alberto Cappellaro}
\email{cappellaro@pd.infn.it}
\affiliation{Dipartimento di Fisica e Astronomia “Galileo Galilei” and CNISM,
Universit\`a di Padova, via Marzolo 8, 35131 Padova, Italy}
\author{Luca Salasnich}
\email{luca.salasnich@unipd.it}
\affiliation{Dipartimento di Fisica e Astronomia “Galileo Galilei” and CNISM,
Universit\`a di Padova, via Marzolo 8, 35131 Padova, Italy}
\affiliation{CNR-INO, via Nello Carrara, 1 - 50019 Sesto Fiorentino, Italy}
\author{Tommaso Macr\`i}
\email{macri@fisica.ufrn.br}
\affiliation{Departamento de F\'isica Te\'orica e Experimental, Universidade Federal do Rio Grande do Norte, 
and International Institute of Physics, Natal-RN, Brazil}
\date{\today{}}

\begin{abstract}
We systematically investigate the zero temperature phase diagram 
of bosons interacting via dipolar interactions in three dimensions in free space 
via path integral Monte Carlo simulations 
 with few hundreds of particles 
and periodic boundary conditions
based on the worm algorithm.
Upon increasing the strength of the dipolar interaction and at sufficiently high densities we find a wide 
region where filaments are stabilized along the direction of the external field. 
Most interestingly by computing the superfluid fraction we conclude 
that superfluidity is anisotropic and is greatly suppressed along the
orthogonal plane. 
Finally we perform simulations 
at finite temperature confirming the stability of filaments against thermal fluctuations and provide an estimate
of the superfluid fraction in the weak coupling limit in the framework of the Landau two-fluid model.
\end{abstract}
\pacs{05.30.-d, 03.75.Hh, 03.75.Kk, 03.75.Nt, 67.85.De, 67.85.Bc, 67.80.K-}

\maketitle

Superfluidity is an amazing phenomenon of quantum mechanical origin that manifests itself macroscopically 
as frictionless flow and lack of response to rotation for small enough angular velocity \cite{pitaevskii2003,pethick2002}.
Several experimental platforms have been used to investigate
quantum matter in the superfluid regime \cite{leggett2006}. Among them,
ultracold gases realize a very clean and controllable many-body playground that
permit the observation of quantum properties  
with unprecedented precision \cite{bloch2008}.
In these systems superfluidity has been 
observed both with bosonic as well as with fermionic atoms
\cite{verney2015,tey2013,pitaevskii2015,hou2013}.
The superfluid fraction, the ratio of the superfluid density to the total density of the system,
has been recently measured in two-component Fermi gas 
interacting via strong contact potentials
\cite{pitaevskii2015,salasnich2010,sidorenkov2013}.

An even richer phenomenology appears when long-range interactions are present. 
The non-local character of the interparticle potential may induce
instabilities of the density that lead to a spontaneous breaking of translational symmetry.
A primary example is the long sought supersolid state, where superfluidity is accompanied
by a crystalline order \cite{boninsegni2012,cinti2014,leonard2017,li2017}.
Recent groundbreaking experiments with dipolar condensates demonstrated the
existence of dense bosonic droplets in trapped configurations \cite{kadau2016,ferrier2016,chomaz2016,ferrier2016-jpb}
and in free space \cite{schmitt2016}.
Beyond mean-field effects \cite{wachtler2016,bisset2016,baillie2016} and three-body interactions \cite{bisset2015} have been invoked 
as the main mechanisms responsible for the stability of these clusters. 
Large scale simulations based on a non-local non-linear Schr\"odinger equation 
have shown very good agreement with the density distribution 
and the excitation spectra observed in laboratory. 
Latest experiments paved the way for the search of phase coherence of droplets,
demonstrating interference pattern via expansion dynamics of the condensate. 
The presence of fringes showed that each droplet is individually phase coherent and thus superfluid,
leaving yet unresolved the question of global phase coherence of the system \cite{ferrier2016}.

In this Letter we report path-integral Monte Carlo (PIMC) results 
for the low temperature properties of a  finite size system of dipolar
bosons in three dimensional free space.
Recent works
showed with similar methods the existence of a window in parameter space leading to
stable self-bound configurations of a single \cite{saito2016} 
or several droplets in a regular arrangement in trapped configurations \cite{macia2016}. 
Yet, no work has investigated the superfluid character of these structures.
Here we address the issue of superfluidity for a wide range of the strength of the dipolar
interaction and density as well as their stability against finite temperature fluctuations.
We carry out large scale simulations, based on a continuous-space worm algorithm 
which allows for an efficient and reliable determination of the superfluid density 
\cite{boninsegni2006,boninsegni2006-pre}. 
We observe an anisotropic character of the superfluid fraction in the inhomogeneous regime of the filaments. 
Our results are consistent with the absence of supersolidity in these systems.
\begin{figure}[t!]
\centering
\includegraphics[width=0.9\columnwidth]{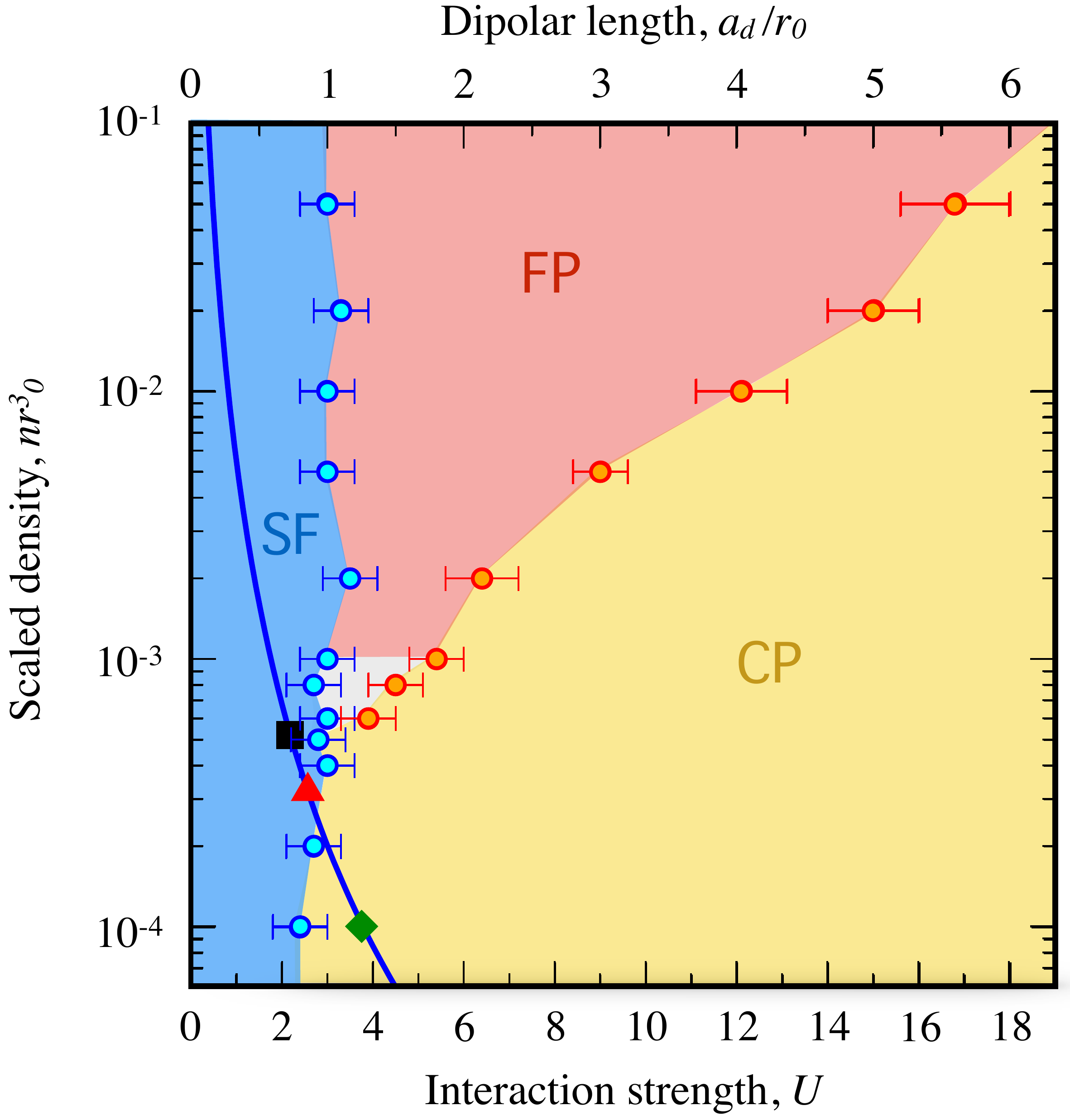}
\caption{Zero-temperature phases of dipolar bosons in three dimensions in free space: 
Superfluid (SF), superfluid filaments (FP) and cluster phases (CP) for varying 
dimensionless interaction strength $U$ (or equivalently dipolar length $a_d/r_0$) and scaled density $n r_0^3$. 
For $a_d/r_0\lesssim 0.9$ the blue SF region corresponds to a uniform superfluid, which 
agrees with mean-field prediction that predicts stability for $a_d/a_s<1$.  
FP is encountered for high densities $nr_0^3\gtrsim 5\cdot 10^{-4}$ and for $a_d/r_0\gtrsim1$.
For small densities $n r_0^3\lesssim 5 \cdot 10^{-4}$ the system
breaks up into tiny clusters. 
Phase boundaries in the grey region are not resolved.
Blue line corresponds to $n=5\cdot 10^{20}\,$m$^{-3}$ and $a_d=130\,a_0$. 
Black/Red dots are experimentally measured background scattering lengths 
$a_s=122\,a_0$ ($^{162}$Dy) / $a_s=92\,a_0$ ($^{164}$Dy). 
Green point at $a_s=29\, a_0$ \cite{nota3}.
}
\label{fig1}
\end{figure}

The Hamiltonian of an ensemble of $N$ interacting identical bosons is:
\begin{equation} \label{ham}
\hat H = -\sum_{i=1}^N \frac{\hbar^2}{2m}\nabla_i^2 + 
 \sum_{i<j}^N V(\mathbf{r}_i-\mathbf{r}_j),
\end{equation}
where $m$ is the particle mass, $\mathbf{r}_i$ is the position of the $i$-th dipole.
We model the interparticle potential $V(\bf{r})$ by a short-range hard-core with cutoff $r_0$
and an anisotropic long-range dipolar potential:
\begin{equation} \label{potential}
V(\mathbf{r})=
\left\{
\begin{array}{cc}
\frac{C_{dd}}{4\pi}\frac{1-3\cos^2\theta}{r^3} & \text{if}~r\ge r_0, \\
\infty & \text{if}~r<r_0,
\end{array}
\right.
\end{equation}
where $C_{dd}/4\pi$ is the coupling constant and the angle $\theta$ denotes the angle between the vector $\bf{r}$ and the 
$z$-axis.
Here the units of length and energy are 
$r_0$ and $\hbar^2/m r_0^2$, respectively. 
Therefore the zero-temperature physics is controlled by the dimensionless 
interaction strength $U=mC_{dd}/4\pi\hbar^2 r_0$ and the dimensionless density
$n r_0^3$ only \cite{nota1}.
Eq.(\ref{ham}) applies both to vertically aligned dipoles interacting via magnetic
or electric dipole moments \cite{lahaye2009}. 
The hard-core with radius $r_0$ 
removes the unphysical $1/r^3$ attraction of dipoles
in head-to-tail configurations at small distances. 
Finite dipolar potentials affect the scattering length $a_s$ associated to 
the contact potential into a non-trivial 
relation of $C_{dd}$
\cite{bortolotti2006,ronen2006,jachymski2016}.
This turns out to be crucial to account for the stability properties of a 
dipolar gas in generic (beyond) mean-field approaches. 
For example a standard Bogoliubov calculation determines
the stability of a homogeneous condensate when
$\epsilon_d\equiv a_d/a_s<1$ is satisfied, where
$a_{d} = \frac{C_{dd}m}{12 \pi \hbar^2}$ is the dipolar length associated to Eq.(\ref{potential})
\cite{lahaye2009,baranov2012}.

We determine 
the equilibrium properties of
Eq.~\eqref{ham} in a wide range of scaled average density $n r_0^3$ (different from 
experimentally measured peak density) and interaction strength $U$.
We work with $N$ atoms in a cubic box of linear dimension $L$ and periodic boundary conditions.
At fixed density $n=N/L^3$, with $N=100-400$, we verified that the phase diagram
does not change. Thus, our finite-size results give a reasonable estimate of thermodynamic limit.
\cite{nota2}. 
From these simulations we obtain, for example, density profiles, and radial correlation functions,
as well as the superfluid fraction \cite{svistunov2015} at finite temperatures.
The  phases were obtained by extrapolating to the 
limit of zero temperature, 
lowering the temperature until observables, 
such as the total energy, superfluid fraction $f_s$
and radial correlations did not change on further decrease of $T$.
These observables 
were then analyzed to construct the phase diagram in Fig.\ref{fig1}.
\begin{figure*}[t!]
\centering
\includegraphics[width=2.0\columnwidth]{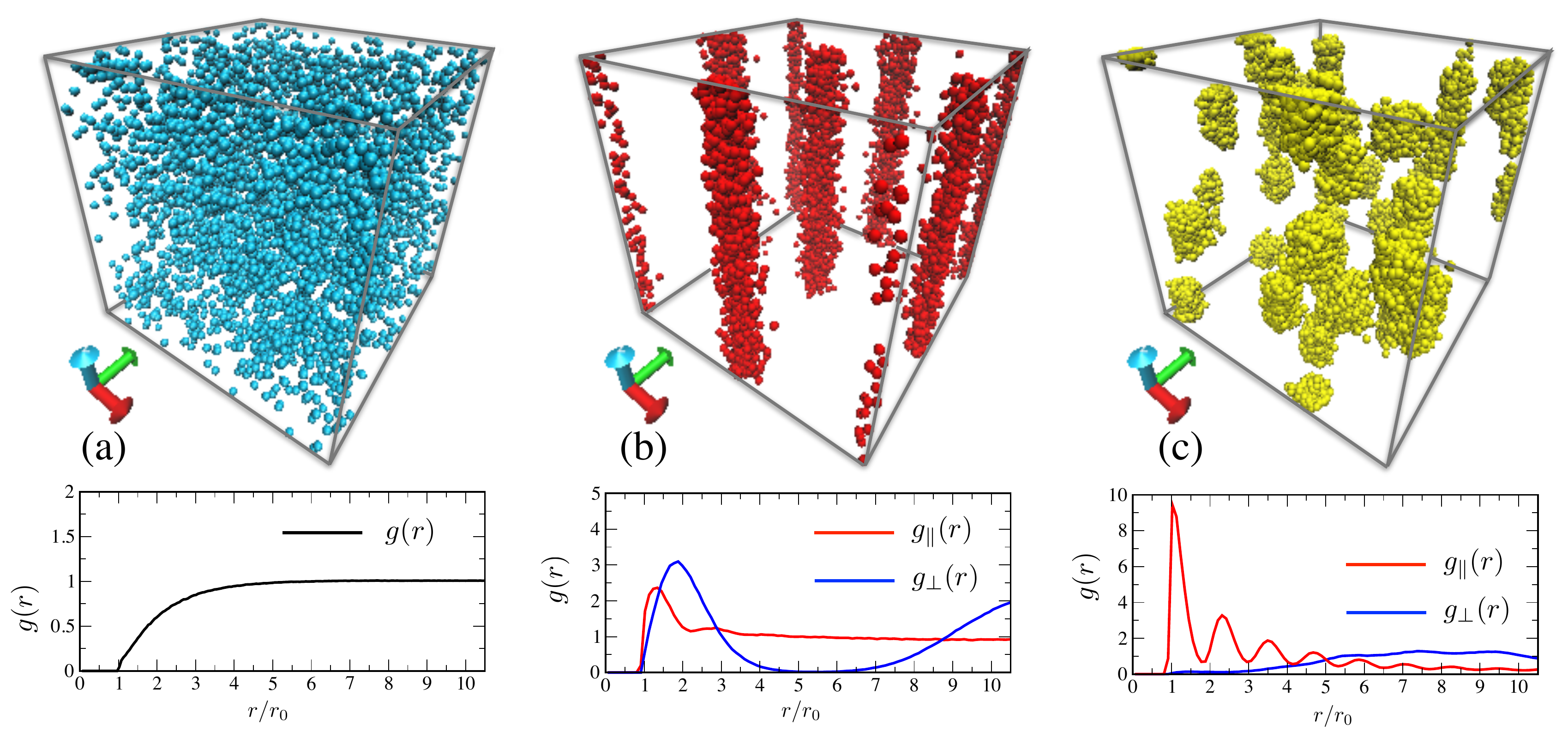}
\caption{
Characterization of the phases at $n r_0^3=10^{-2}$.
Upper panels. PIMC density distribution at different strengths of dipolar interaction: 
(a) SF at $a_d = 0.6 r_0$, 
(b) superfluid FP at $a_d = 2.6 r_0$ and (c) CP at $a_d = 6.0 r_0$.
Lower panels. (a) radial correlation functions $g(r)$ in the SF.
(b) and (c) radial correlation functions $g_{\parallel}(r)$ along the vertical direction (red) and $g_\perp(r)$
along the orthogonal plane (blue). 
$g_{\parallel}(r)$ has a fluid-like profile in the FP. In the CP $g_{\parallel}(r)$
a high peak due to strong attractive interactions appears at distances $r\gtrsim r_0$.
$g_{\perp}(r)$ in the FP shows a strong suppression in between two filaments whereas in the CP it flattens  
at intermediate distances. 
Simulations are done with $100$ particles and $500$ slices. 
Correlation functions are averaged over $100$ configurations.
}
\label{fig2}
\end{figure*}
For small $U$ the system is in a superfluid phase (SF) with unitary $f_s$. 
For low densities $n\,r_0^3\lesssim10^{-3}$
the SF extends up to $U\approx 2.1$ which agrees with the mean-field phase boundary
predicted by standard Bogoliubov analysis ($\epsilon_d=1$) 
for the dipolar potential and a contact interaction
\cite{nota3}. 
Increasing the interaction strength the system enters into a cluster phase (CP) 
with vanishing superfluidity and characterized by droplet structures with few particles.
For higher densities $n r_0^3>6\cdot10^{-4}$,
crossing the SF phase boundary, 
we encounter a phase characterized by elongated 
filaments (FP) with anisotropic superfluid fraction. 
We notice that this FP extends from small positive induced contact potential ($\epsilon_d \gtrsim 1$) 
which corresponds to the experimentally relevant regime of 
\cite{kadau2016,nota3,yi2000,yi2001,tang2015,maier2015,burdick2016,chaikin1995},
to the strongly coupled limit of large dipolar interactions and large densities. 
For intermediate densities $6\cdot 10^{-4} \lesssim n\, r_0^3 \lesssim 1\cdot 10^{-3}$ 
the phase boundaries are not resolved.
\begin{figure}[h!]
\centering
\includegraphics[width=0.9\columnwidth]{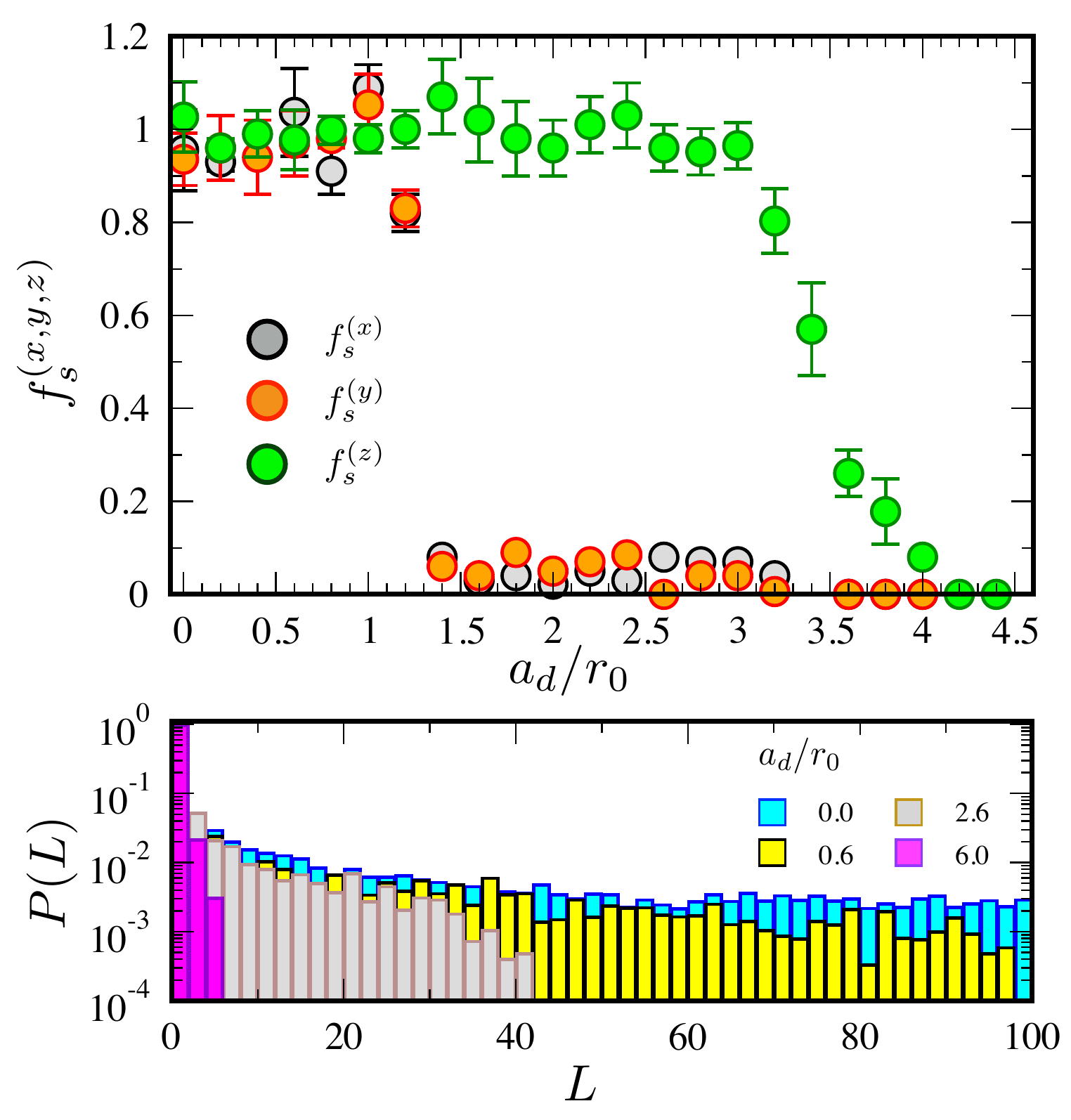}
\caption{Superfluid fraction $f_s$ across the transition from superfluid to filament transition at high densities.
Upper panel. $f_s$ as a function of the dipolar length along the vertical direction and the orthogonal direction
for $n r_0^3=10^{-2}$.
Dipoles and filaments are aligned along the $z$-axis.
In the SF the superfluid fraction converges to $f_s=1$ isotropically. In the filaments we observe $f_s=1$ along the
vertical axis and vanishing on the orthogonal plane.
Error bars are statistical uncertainties.
Lower panel. Relative frequency of permutation cycles of length $L$ again 
for $n r_0^3=10^{-2}$ and $a_d=0,0.6,2.8,6.0\, r_0$ in the the SF (blue and yellow), 
FP (grey) and CP (purple).
}
\label{fig3}
\end{figure}

Representative PIMC configurations with $100$ particles and $500$ 
imaginary-time slices ($n\, r_0^3=10^{-2}$) of the density distribution are shown in Fig.\ref{fig2}. 
In the SF (Fig.2a) particles delocalize into a phase with uniform density 
displaying flat radial distribution functions, $g(r)$, 
typical of a fluid phase of particles interacting via hard-core
potentials.
In Fig.\ref{fig2}b a typical configuration of the FP is illustrated where filaments elongate 
vertically. 
Finally in the CP several small clusters form, slightly elongated along the vertical direction,
signaling a fragmentation of the filaments for large interactions.
In the non-homogeneous FP and CP we analyze $g(r)$
along two orthogonal directions. 
The function $g_\parallel(r)$ (red line in Fig.\ref{fig2}) is computed along the vertical axis parallel to the direction of the filaments. 
In the FP $g_\parallel(r)$ has a liquid-like profile with a peak $r\gtrsim r_0$ as a consequence of the attractive 
part of the dipolar interaction and the large density along the $z$-axis. In the CP the first peak of $g_\parallel(r)$
is higher and the function oscillates several times before vanishing for large distances as a consequence of the
finite size of the clusters. 
The correlation $g_\perp(r)$ (blue line) is evaluated along the horizontal plane, perpendicular to the direction of the filaments. 
In the FP, it is strongly suppressed 
in the region between two filaments.
In the CP $g_\perp(r)$ flattens at intermediate distances 
displaying an irregular arrangement of the small dipolar droplets
along the horizontal plane.

To fully characterize the genuine quantum properties of the FP we compute the superfluid fraction 
$f_{s}^{(i)}$ as a function of the interaction strength and temperature along three orthogonal directions $i=\,x,\,y,z$, 
yielding 
$f_{s}^{(i)}=\frac{m k_BT}{\hbar^2\,n}\langle w_{i}^2\rangle$,
where 
$\langle \ldots \rangle$ stands for the  thermal average of the 
winding number estimator $w_i$
\cite{pollock1987,ceperley1995}.
Fig.\ref{fig3} depicts the superfluid fraction $f_s^{(i)}$ calculated  
varying the dipolar length $a_d$ for $n r_0^3=10^{-2}$. 
In the SF the superfluid density is uniform and unitary ($f_{s}^{(i)}=1$).
Crossing the SF-FP phase boundary $f_{s}^{(i)}$ displays a strong anisotropy. 
The superfluid fraction is unitary along the vertical direction
and it is greatly suppressed along the orthogonal $x-y$ plane, showing only a marginal 
finite-size effect contribution to $w_x$ and $w_y$, respectively. 
This result suggests that each filament is phase coherent, but globally the system is not.
This suppression of superfluidity along two directions resembles the formation of a sliding phase
in weakly-coupled superfluid layers in the thermodynamic limit \cite{ohern99,pekker10,laflorencie12,vayl17}.
Finally entering in the CP $f_{s}^{(i)}=0$ uniformly as expected from the fragmentation of the 
filaments in the configurations.
The lower panel of Fig.\ref{fig3} reports the relative frequency of exchange cycles involving 
a particle number $1\le L \le N$ \cite{jain2011}. 
In the SF we observe long permutation cycles ($\sim N$).
In the FP (CP) cycles reduce to few tens of (few) particles in support of the previous 
observation that superfluidity originates within each filament (locally within each cluster).

\begin{figure}[t!]
\centering
\includegraphics[width=0.9\columnwidth]{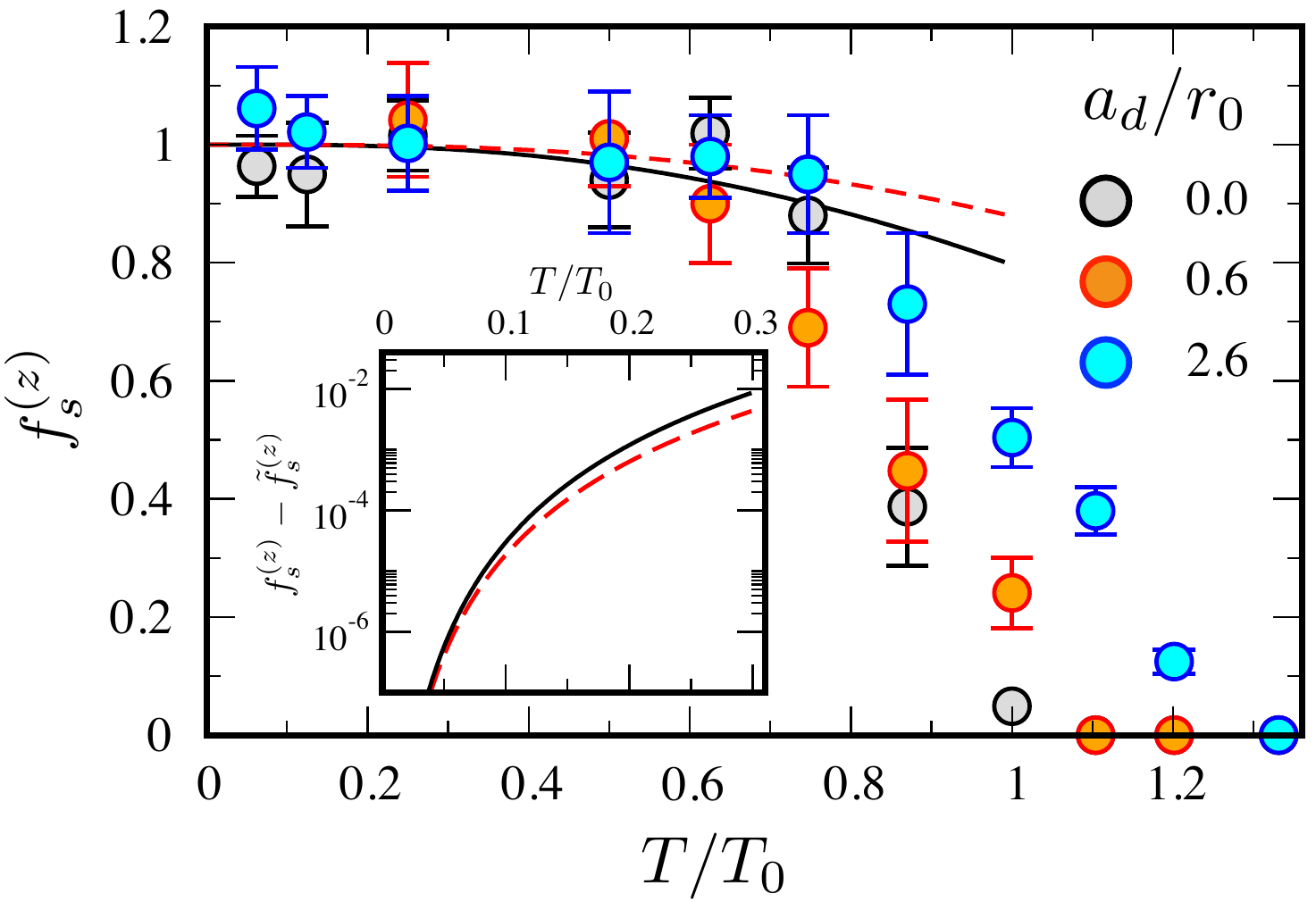}
\caption{Superfluid fraction $f_s^{(z)}$  
as a function of the scaled temperature $T/T_0$
at $n r_0^3=10^{-2}$ for three values of the dipolar lengths:
a) SF with $a_d=0$ (hard-core bosons), 
b) SF with $a_d=0.6 r_0$, c) FP with $a_d =2.6\, r_0$.
Lines refer to the analytical prediction of the temperature dependence of the superfluid fraction
at $n r_0^3=10^{-2}$ from Eq.(\ref{fs}) at $a_d=0 $ (black continuous line) 
and $a_d=0.6\, r_0$ (red dashed line).
Inset. Low temperature limit of $f_s^{(z)}$ from Eq.(\ref{fs}) and Eq.(\ref{fs_low_T}) 
in the SF at $a_d=0 $ and $a_d=0.6\, r_0$.
}
\label{fig4}
\end{figure}

Finally we investigate the stability of superfluid filaments at finite temperatures. 
In Fig.\ref{fig4} we show the superfluid fraction for a system of $N=100$ particles at density 
$n r_0^3=10^{-2}$ as a function of $T/T_0$, $T_0$ being
the critical temperature of an ideal Bose gas 
$k_B T_0 = \frac{2\pi}{\zeta\left(3/2\right)^{2/3}} \left(nr_0^3\right)^{2/3}$
\cite{nota4}.
In the SF we compare PIMC results with
an analytical calculation within the Landau two-fluid model \cite{ghabour2014}
\begin{equation} \label{fs}
f_s^{(z)} = 1 - \frac{\beta \hbar^2}{4\pi^2 m} \int_0^{\infty}dq \int _0^{\pi}d\theta_q q^4 \frac{ \sin\theta_q \cos^2\theta_qe^{\beta E_q}}{(e^{\beta E_q} - 1)^2}
\end{equation}
where $E_\mathbf{q}$=$\sqrt{\frac{\hbar^2q^2}{2m}\left(\frac{\hbar^2q^2}{2m}+2\frac{\mu}{g_0} \tilde V_\text{ps}(\mathbf{q}) \right)}$ 
is the Bogoliubov spectrum. 
The function $\tilde V_\text{ps}(\mathbf{q})=g_0\left(1-\epsilon_d + 3\epsilon_d \cos^2(\theta_q)\right)$ 
is the Fourier
transform of the pseudo-potential
$V_{\text{ps}}({\bf r})$=$g_0\delta({\bf r})+\frac{C_{dd}}{4\pi}\frac{1-3\cos^2\theta}{r^3}$, 
widely used as a regularized form of the interparticle potential (\ref{potential}),
with $g_0$=$4\pi\hbar^2a_s/m$
\cite{baranov2008,lahaye2009,baranov2012}.
The chemical potential $\mu$ includes the beyond mean field contribution of the dipolar interaction:
\begin{equation}
\mu = g_0 n \left(1+ \frac{32}{3\sqrt{\pi}} \sqrt{n a_s^3}\, Q_{5/2}(\epsilon_d)\right),
\end{equation}
where $Q_\alpha(x)$=$ \int_0^1dt\, \left(1- x+3xt^2\right)^\alpha$  \cite{lima2011}.
The plot of the superfluid fraction along the vertical axis from Eq.(\ref{fs}) is shown in Fig.4 
for $a_d=0$ (black line) and $a_d=0.6\,r_0$ (red dashed line).
In the low temperature regime we can linearize $E_\mathbf{q}$ to determine an 
approximate analytical formula for the superfluid fraction:
\begin{equation} \label{fs_low_T}
\tilde{f}_s^{(z)}=1-\frac{2\pi^2}{45} \frac{1}{(1-\epsilon_d)(1+2\epsilon_d)^{3/2}}\left(\frac{m}{\hbar^2} \right)^{3/2}\frac{(k_B T)^4}{n\,\mu^{5/2}},
\end{equation}
that works well up to $T\lesssim0.3\, T_0$. In the inset of Fig.\ref{fig4} we show the difference between the 
superfluid fractions of Eq.(\ref{fs}) and the analytical result Eq.(\ref{fs_low_T}) for the same values of $a_d$ 
as in the main figure.
For larger dipolar length $a_d=2.6\, r_0$, in the FP, the superfluid
fraction along the vertical axis is finite within a large window of temperatures. We conclude then that 
filaments are stable against finite temperature fluctuations and we clearly see that anisotropic 
superfluidity is finite up to temperatures $T \sim$ $0.8\, T_0$. We also verified that the orthogonal 
components of the superfluid fraction are vanishing in this regime. 

Our results then suggest that the observed interference pattern
\cite{kadau2016} is a consequence of {\emph{local}} phase coherence
(within each droplet) and not a {\emph{global}} one. 
As discussed above the FP is characterized by an anisotropic superfluid density
that, interestingly, might be measured via the second sound along the orthogonal directions 
as done with strongly interacting Fermi gases \cite{verney2015,tey2013,pitaevskii2015,hou2013}.

In this Letter we studied the many-body phases of an ensemble of bosons interacting via dipole-dipole interactions
in three dimensional free space and investigated the superfluid behavior of dipolar filaments. 
We spanned a wide range of the parameter space 
confirming the existence of an extended phase
of filaments with unitary superfluidity along the vertical direction
and vanishing otherwise. 
Our results therefore theoretically support recent experimental findings
about the stability of droplets in free space \cite{schmitt2016} and  the presence of local
phase coherence \cite{kadau2016} giving rise to interference fringes but exclude global phase coherence of the filaments \cite{nota3}
and therefore a possible supersolid phase.
Finally we confirmed that filaments are stable at finite temperature. 
More refined investigations are needed to determine accurately the melting transition 
of the filaments into a fluid phase.

\begin{acknowledgments}
{\it Acknowledgments}.
We thank R. N. Bisset, M. Boninsegni, S. Giorgini, G. Gori, A. Pelster, T. Pohl, 
G. Shlyapnikov, F. Toigo and M. Troyer
for useful discussions and I. Ferrier-Barbut
for helpful correspondence.
T.M. acknowledges CNPq for support through Bolsa de produtividade
em Pesquisa n. 311079/2015-6 and thanks NITheP for the hospitality where part of the work was done. 
F.C. acknowledges the hospitality of the
MPI-PKS in Dresden where part of the work was carried out.
\end{acknowledgments}

\pagebreak
\widetext
\begin{center}
	\textbf{\large Supplemental Materials: Superfluid filaments of dipolar bosons}
\end{center}
\setcounter{equation}{0}
\setcounter{figure}{0}
\setcounter{table}{0}
\setcounter{page}{1}
\makeatletter
\renewcommand{\theequation}{S\arabic{equation}}
\renewcommand{\thefigure}{S\arabic{figure}}
\renewcommand{\bibnumfmt}[1]{[S#1]}
\renewcommand{\citenumfont}[1]{S#1}

\subsection{Density profile in the Filament Phase}
\begin{figure}[h!]
	\centering
	\includegraphics[width=0.5\columnwidth]{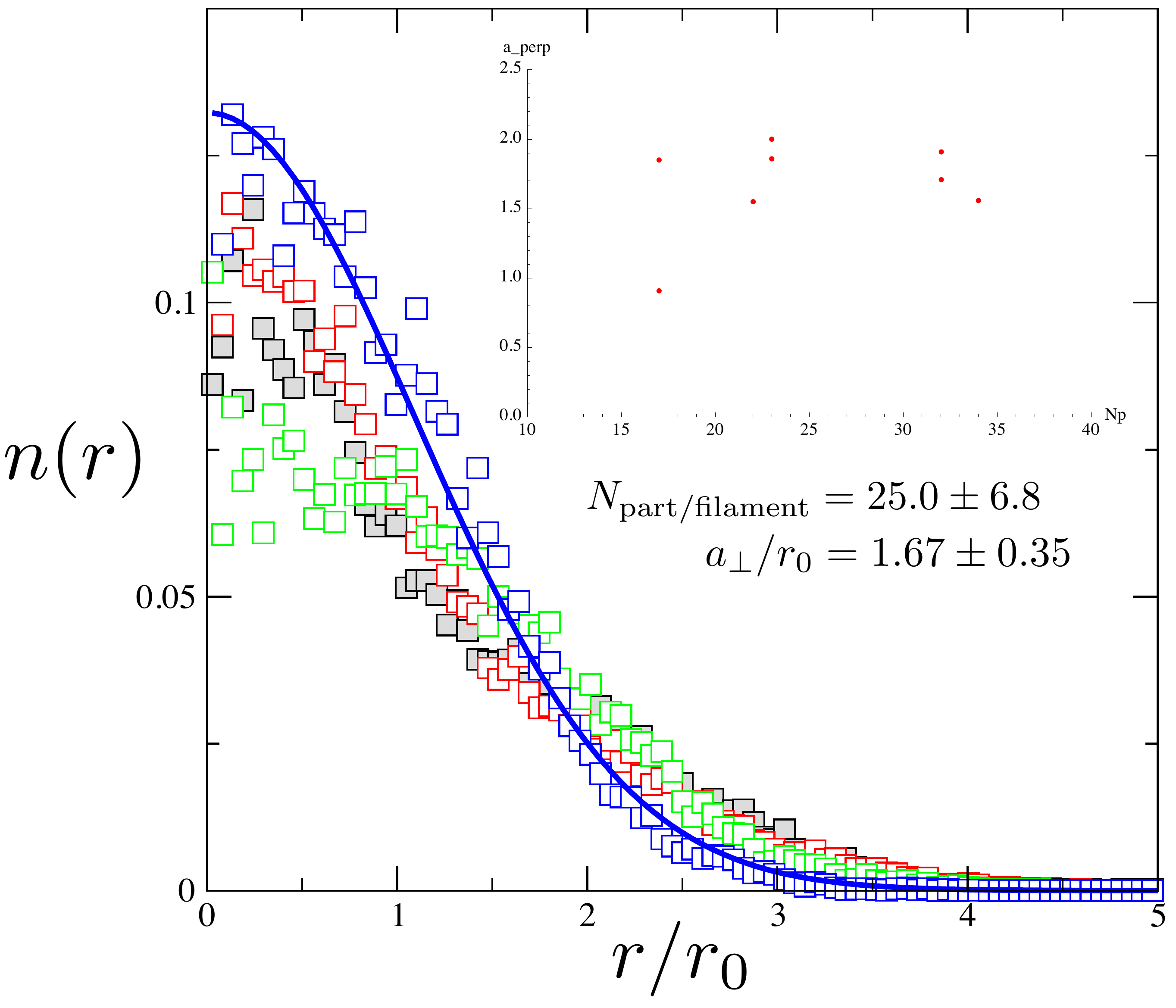}
	\caption{Particle density within a filament as a function of $r/r_0$. For this filament the average particle number is $N=25$.}
	\label{figS1}
\end{figure}

In Fig.\ref{figS1} we analyze the density on few filaments in the FP at 
$n\,r_0^3=10^{-2}$. We fit the density with a gaussian and 
extract the average number of particles within each filament. The inset shows the width 
of the gaussian in units of $r_0$ as a function of the particle number. The estimate of the average 
inter-particle distance with a gaussian wave packet in the radial direction and a cylinder of length 
$L$ along $z$ gives
\begin{equation}
	\left< r \right> \approx \frac{1}{n^{1/3}}= \left(\frac{\pi a_\perp^2 L}{N}\right)^{1/3} = 2.1\, r_0
\end{equation}
for the data of Fig.S1 below. This result is also in agreement with the calculation 
of the radial correlation function $g_\perp(r)$ of Fig.2b (bottom) which is peaked at $r \approx 2 r_0$.

\subsection{Relation among scattering length, dipolar length and cutoff radius and universality}

\begin{figure*}[h!]
	\centering
	\includegraphics[width=0.5\columnwidth]{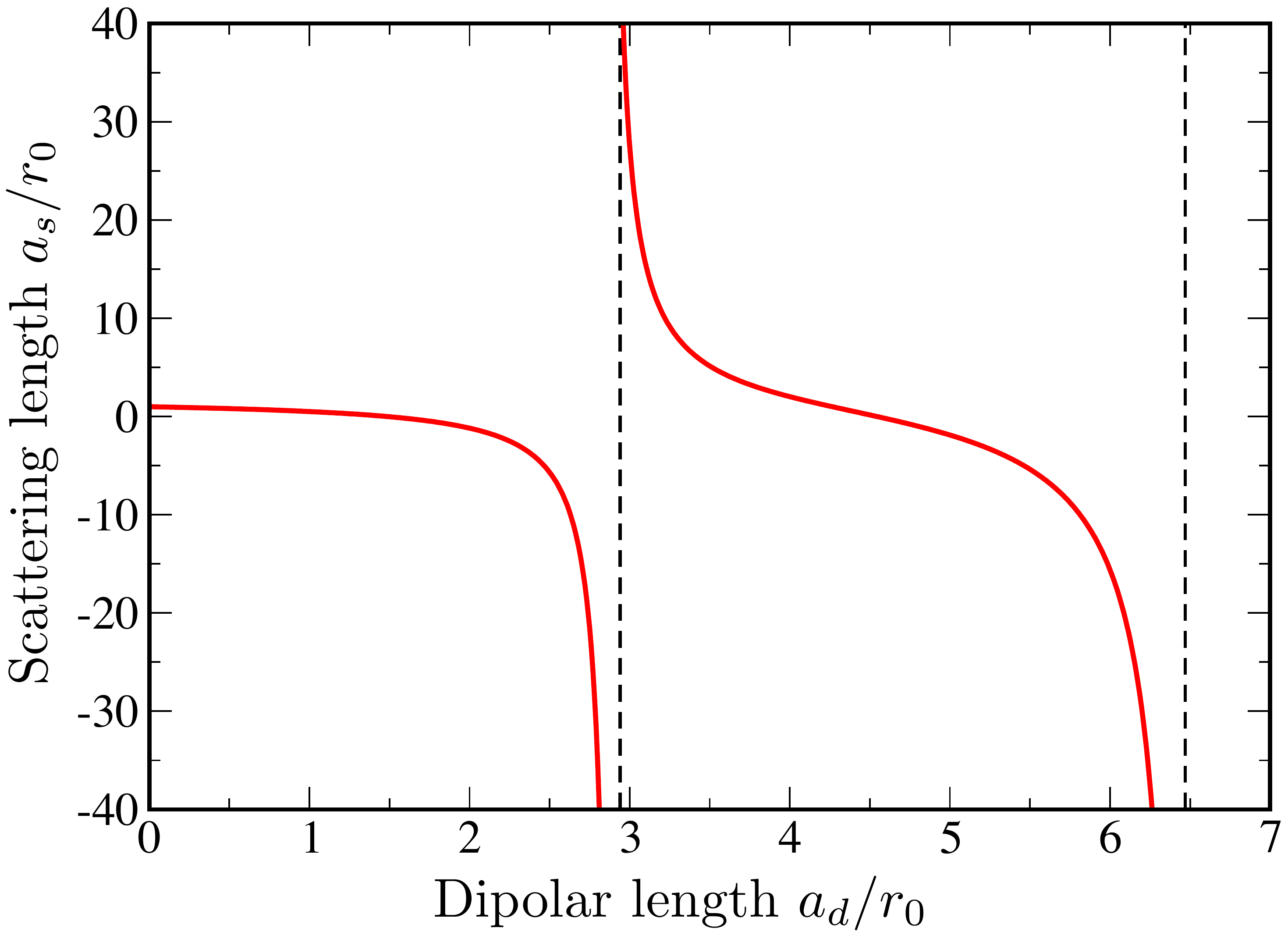}
	\caption{
		Relation between scattering length $a_s$ and dipolar length $a_d$ in units of $r_0$. Data are taken from \cite{ronen2006}.
	}
	\label{figS2}
\end{figure*}

\begin{figure}[h!]
	\centering
	\includegraphics[width=0.7\columnwidth]{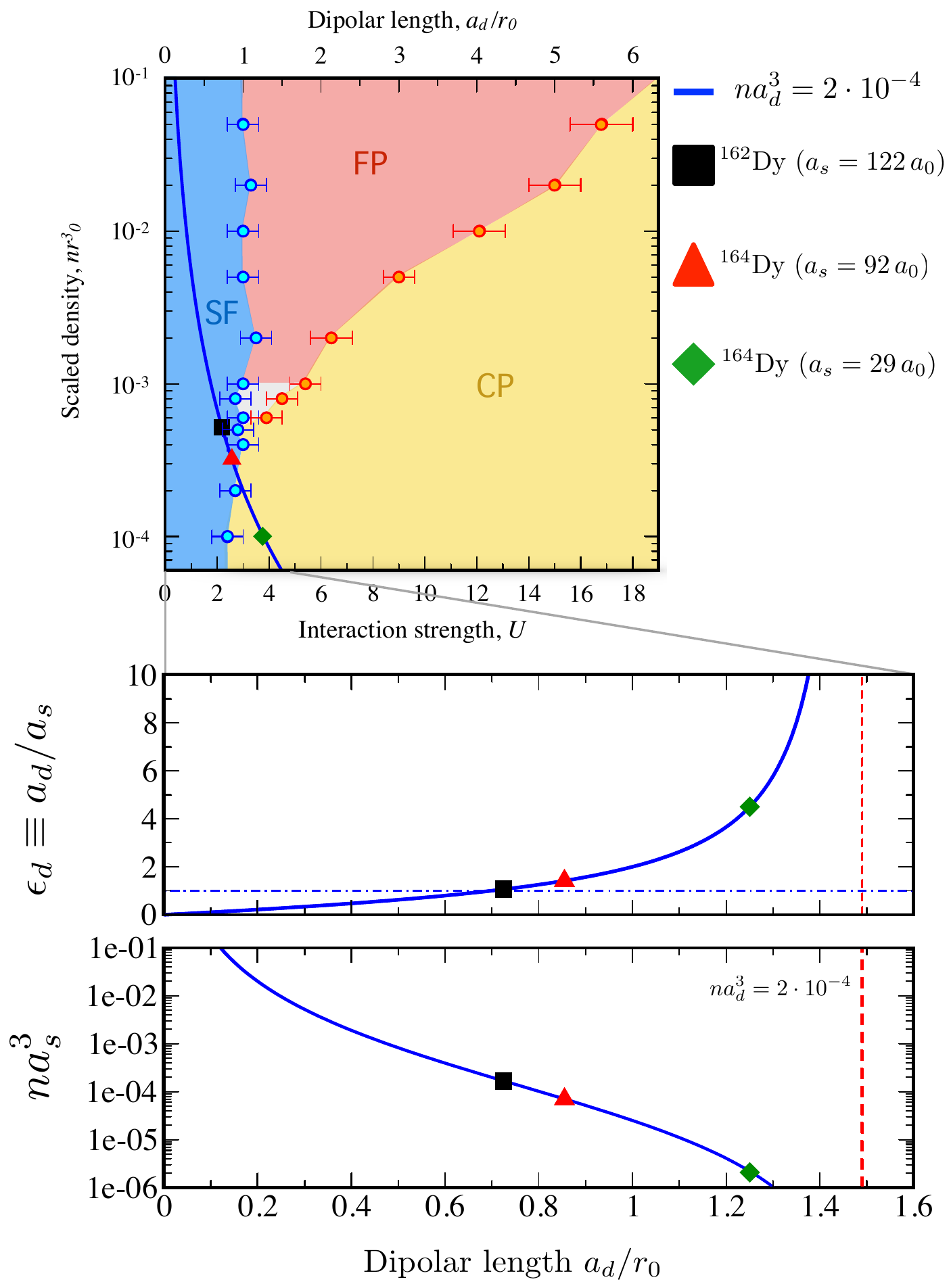}
	\caption{$^{162}$Dy and $^{164}$Dy phase line for $n\cdot a_d^3 = 2\cdot 10^{-4}$. 
		(Top) Upon fixing the average density to $n=5\cdot 10^{20}$ m$^{-3}$ and the dipolar length to $a_d = 130\,a_0$ 
		we find that the condition $n\cdot a_d^3 = 2\cdot 10^{-4}$ 
		is parametrized by the solid blue line in the phase diagram above.  
		(Bottom) Ratio $\epsilon_d$ between the dipolar length $a_d$ and the 
		scattering length $a_s$ computed from Fig.\ref{figS2} above. Below we plot the parameter $n\, a_s^3$ 
		along the blue line in the phase diagram.
		The red dashed line represents the line for 
		which the scattering length as vanishes.
		Black square and red triangle correspond to experimental parameters for Dy with their background scattering length ($122\,a_0$
		and $92\, a_0$ respectively).
		Green diamond is the point for Dy with $a_s=29\,a_0$ for which $n\,r_0^3=10^{-4}$ in the CP.}
	\label{figS3}
\end{figure}

The relation between the dipolar length and the scattering length for the potential of Eq.(2) 
of the text is discussed in detail in \cite{bortolotti2006,ronen2006}. 
In Fig.\ref{figS2} below we show the relation between $a_s$ and $a_d$ in units of $r_0$.
At $a_d/r_0 \approx 2.9$ and $a_d/r_0\approx 6.5$
there are resonances at which the scattering length $a_s$ diverges. 
From Fig.\ref{figS2} one can extract the value of $\epsilon_d = a_d/a_s$ as a function of the corresponding 
ratios $a_s/r_0$ and $a_d/r_0$ (see below Fig.\ref{figS3}). 

\subsubsection{Universality of the results of the phase diagram obtained with the model potential}
The cutoff radius $r_0$ of the model potential in Eq.2 in the main text 
is used to enable simulations within our QMC numerical scheme. 
In \cite{bortolotti2006,ronen2006} the authors numerically 
compute the full low-energy scattering amplitude for the same model potential
and compare it with the one obtained from the first Born approximation within 
the pseudo-potential of Yi and Yu \cite{yi2000,yi2001}
\begin{equation}
	V_{\text{eff}}(\mathbf{r}) = \frac{4\pi \hbar^2 a_s(a_d)}{m} \delta(\mathbf{r}) + \frac{C_{dd}}{4\pi}\frac{1-3\cos^2\theta}{r^3}.
\end{equation}
Upon varying the ratio $a_d/r_0$ they find a remarkably good matching of the
scattering amplitudes obtained with the two approaches
in the very low energy limit, once including an explicit dependence of the scattering length $a_s(a_d)$ 
on the dipolar length $a_d$. 
In this respect the results obtained in our phase diagram lie in the universal regime, 
since only $a_s$ and $a_d$ are relevant parameters. 

It is worth mentioning that more realistic model potentials (with Van der Walls short range potentials) 
were recently considered in \cite{jachymski2016} for the description of Dysprosium 
two-body collisions.
These potentials eventually lead to the same effective potential 
of Yi and Yu above, with a dipolar length depending on the short range scattering length.

\subsection{Phase diagram and comparison with experimental parameters of $^{162}$Dy and $^{164}$Dy}
\begin{figure}[h!]
	\centering
	\includegraphics[width=0.7\columnwidth]{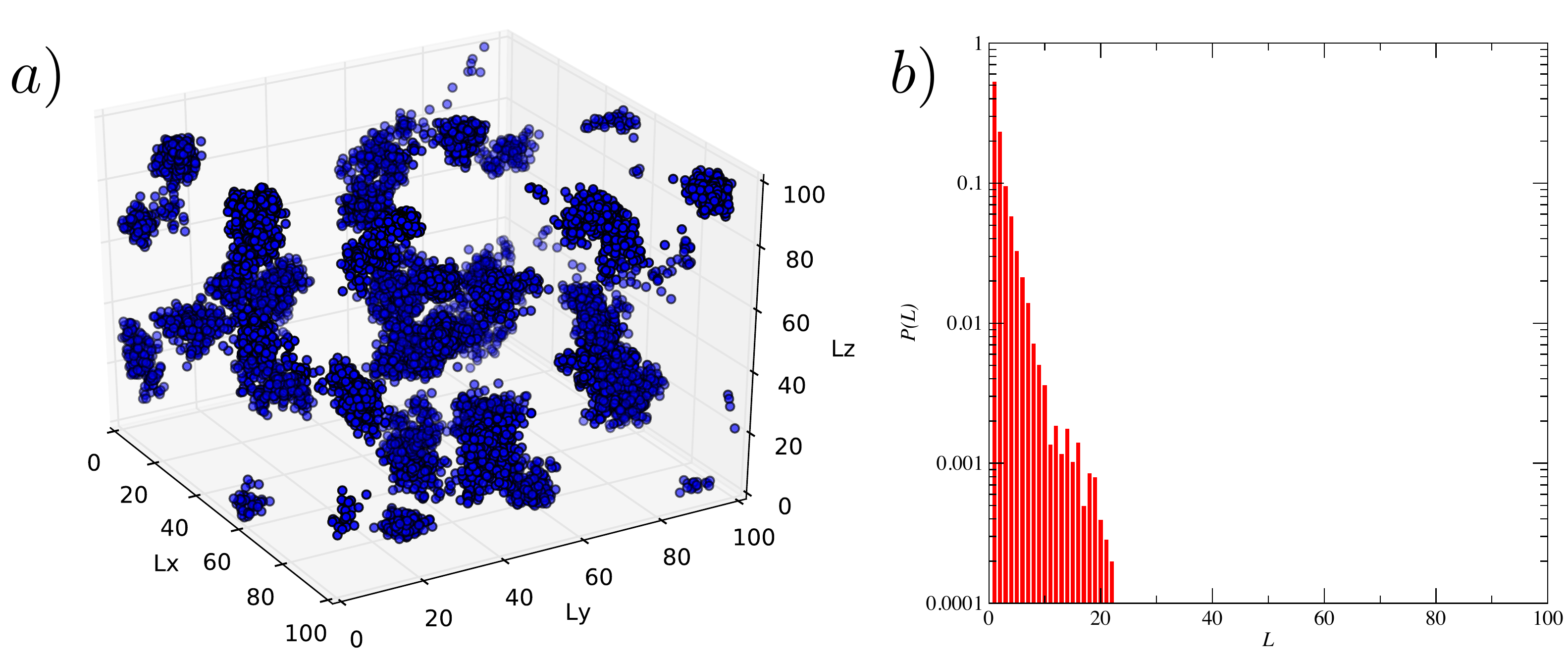}
	\caption{a) Configuration in the cluster phase for $a_d/r_0=1.2$ and $n\, r_0^3 = 10^{-4}$. 
		The simulation is done along the blue curve in the phase diagram.
		b) Permutation cycles for the same parameters in fig.a).}
	\label{figS4}
\end{figure}

We compare our theoretical results with recent experiments with $^{162}$Dy and 
$^{164}$Dy which have dipolar lengths $a_d = 129.2\, a_0$ and $130.8\, a_0$ respectively. 
As an example we analyze the case of fixed average density $n = 5\cdot 10^{20}$ m$^{-3}$ and 
derive the blue curve in our phase diagram corresponding to this condition.
Namely from $n\,a_d^3=2\cdot10^{-4}$, we write:
\begin{equation}
	n\, r_0^3=\frac{n\,a_d^3}{\left(a_d/r_0\right)^3}.
\end{equation}
(For simplicity we fix $a_d=130\,a_0$ for both components since the deviation is not significant).
The background scattering length were recently measured for both isotopes 
\cite{tang2016,maier2015,burdick2016}:
\begin{equation}
	^{162}\text{Dy}=122\, a_0 \ \ \ \ ^{164}\text{Dy}=92\, a_0.
\end{equation} 
Using the parametrization of Fig.\ref{figS2} we are able to identify the two points in the phase diagram.
Notice that the experimental results lie in the transition region (within the error bars) from SF to CP.
In the lower part of the figure we zoom on the relevant part of the phase diagram and compute 
$\epsilon_d=a_d/a_s$ and the parameter $n\, a_s^3$. 
From the calculation of $\epsilon_d$ for the two isotopes we can also compute the value of the 
parameter $r_0$:
\begin{equation}
	^{162}\text{Dy:}\ r_0 = 177\,a_0 \ \ \  ^{164}\text{Dy:}\ r_0 = 154\,a_0.
\end{equation}
In Fig.\ref{figS4}-a we show a configuration in the cluster phase for $a_d=1.25\, r_0$ corresponding to $\epsilon=4.5$ 
($a_s=29\,a_0$) at $n\,r_0^3=10^{-4}$ along the same blue line of the experimental data.
In the right panel Fig.\ref{figS4}-b we plot the permutation cycles for the same parameters, showing permutations
within the cluster, supporting the interpretation that locally particles exchange and quantum droplets
are locally superfluid. 


\subsection{Weakly interacting regime}
\begin{figure}[h!]
	\centering
	\includegraphics[width=0.5\columnwidth]{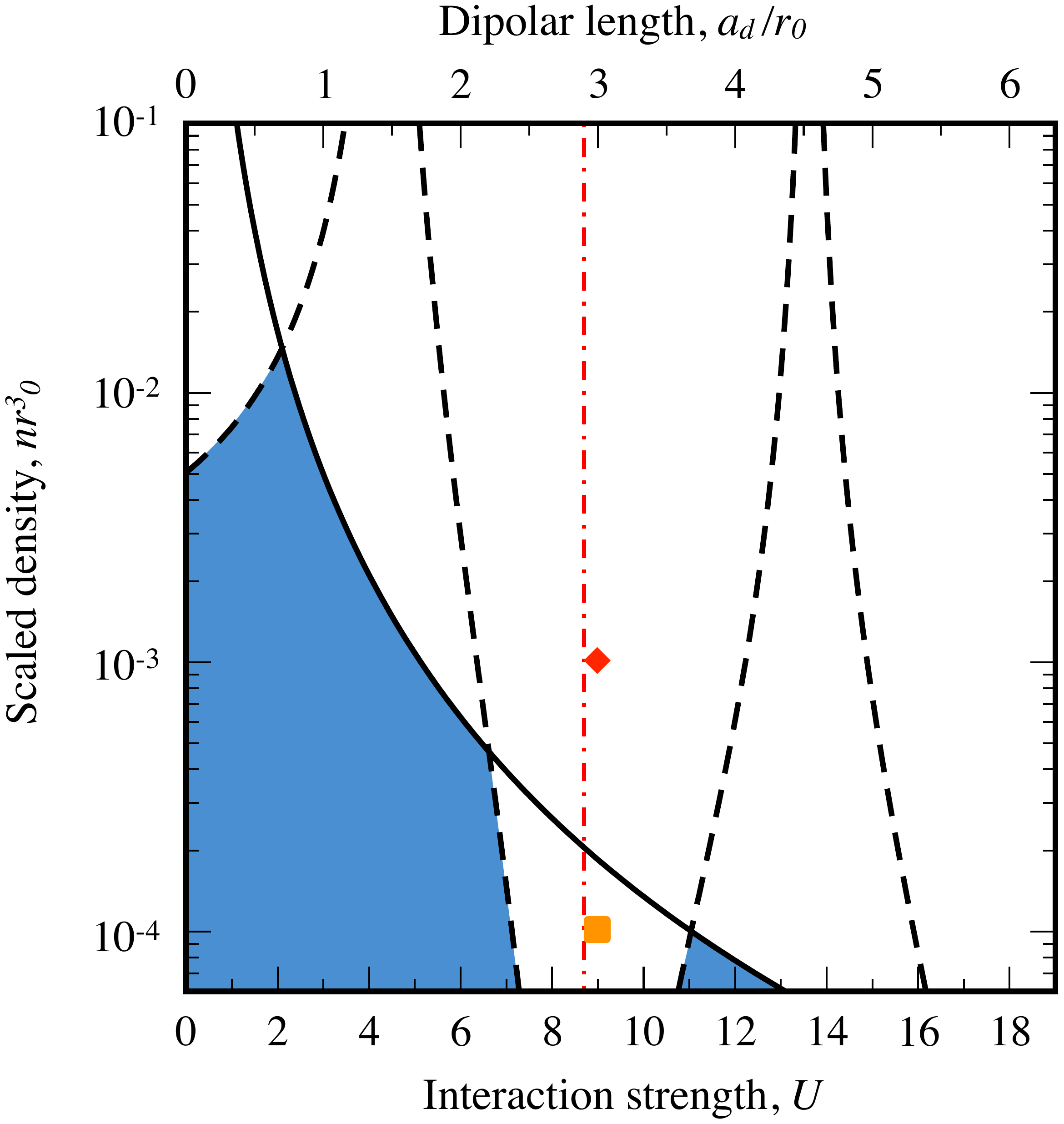}
	\caption{Universality of the model potential used in simulations. In the interval used in the simulations we plot the 
		curves corresponding to $n\,\left|a_s\right|^3=5\cdot 10^{-3}$ (dashed) and $n\,a_d^3=5\cdot 10^{-3}$ (solid). 
		The blue region displays the parameter space where both the inequalities 
		$n\,\left|a_s\right|^3\le 5\cdot 10^{-3}$ and $n\,a_d^3\le 5\cdot 10^{-3}$ 
		are satisfied. The red dashed line shows the position of the resonance where divergence of the scattering length diverges.
		The red diamond at $n\, r_0^3 = 10^{-3}$ and the orange square at $n\, r_0^3 = 10^{-4}$ and $a_d/r_0=3$ correspond to
		configurations of Fig. \ref{figS5}. 
	}
	\label{figS5}
\end{figure}

In our phase diagram we vary $n\,r_0^3$ in the range $6\cdot 10^{-5}-10^{-1}$. 
To answer the question whether the many-body physics of our model
could be described by a \emph{universal weakly interacting theory}, 
we compute $n\,\left|a_s\right|^3$ and $n\,a_d^3$ and verify whether
the conditions:
\begin{equation} 
	n\,\left|a_s\right|^3\ll 1, \ \ \ \ \ n\,a_d^3\ll 1
\end{equation} 
are simultaneously satisfied. In other words, both potential length scales are much smaller than the 
inter-particle distance.
In Fig.\ref{figS5} we plot the curves corresponding to $n\,\left|a_s\right|^3=5\cdot 10^{-3}$ (dashed lines) and 
$n\, a_d^3=5\cdot 10^{-3}$ (solid lines). The shaded region depicts the region where both inequalities:
\begin{equation} \label{ineq}
	n\,\left|a_s\right|^3\le 5\cdot 10^{-3}, \ \ \ \ \ n\,a_d^3\le 5\cdot 10^{-3}
\end{equation} 
hold.
Note that the {\it weakly interacting} regime is the one for small values of the ratio $a_d/r_0$, and that the presence
of a resonance at $a_d/r_0\approx 2.9$ introduces a small region where the inequalities (\ref{ineq}) are both satisfied.
In Fig.\ref{figS5} we show two snapshots of QMC configurations close to resonance at $a_d/r_0= 3$ for $n\, r_0^3 = 10^{-4}$
and $n\, r_0^3 = 10^{-3}$ respectively. In both cases $\epsilon=0.12$, and a mean field theory would predict
a superfluid phase. However as we commented above, this is not necessarily true, since in both cases we are outside
the weakly interacting regime. Indeed both configurations clearly show that the system is in a cluster phase for such
parameters.

\begin{figure}[h!]
	\centering
	\includegraphics[width=0.7\columnwidth]{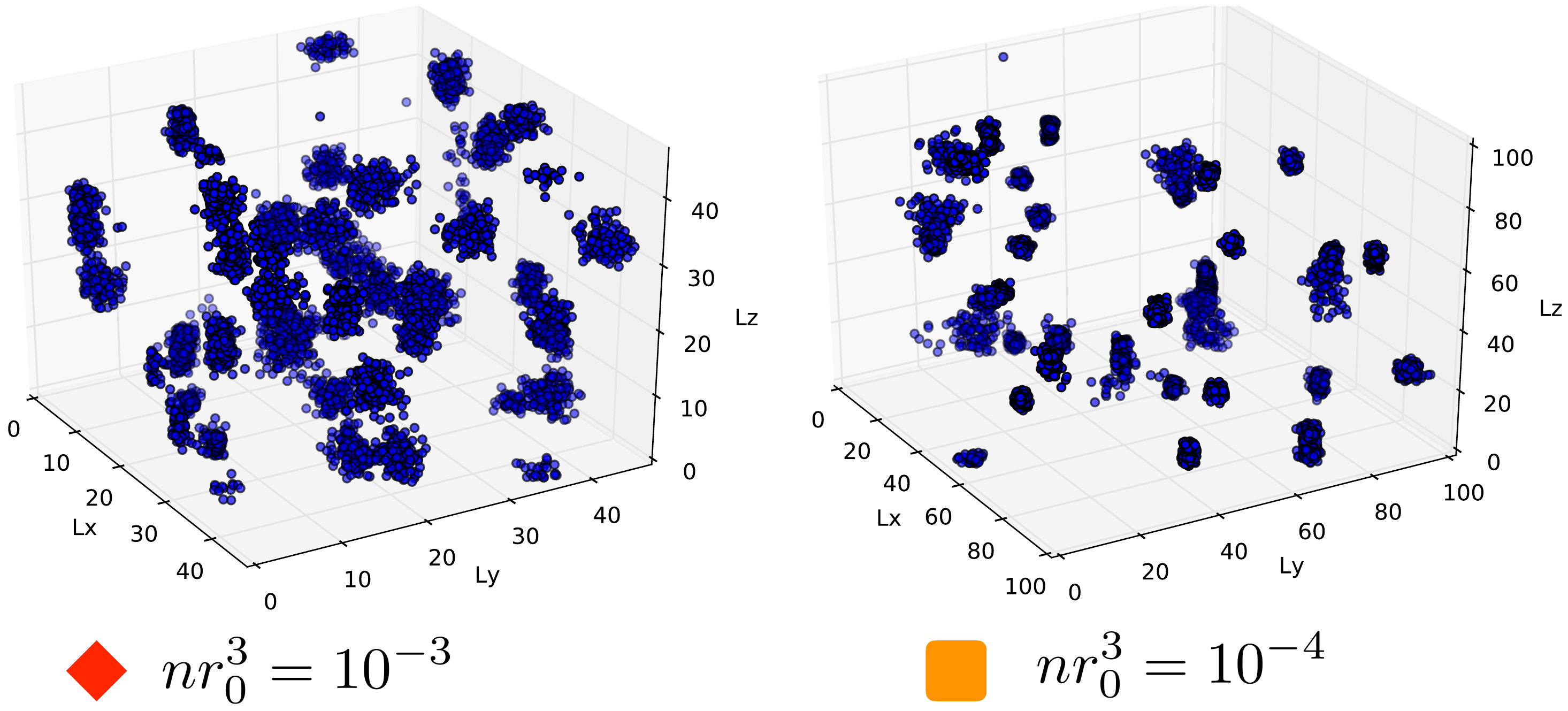}
	\caption{Snapshots of configurations for $N=100$ particles at $a_d/r_0=3$. (Left) $n\, r_0^3 = 10^{-4}$, (right)
		$n\, r_0^3 = 10^{-3}$. For both the scattering length equals $a_s/r_0=26$, therefore $\epsilon=0.12$. 
		Both configurations refer to the points in Fig.S4.
		The system is in the cluster phase and not in a superfluid one contrarily to what is predicted by mean field theory.}
	\label{figS6}
\end{figure}

\subsection{Radial distribution function}
In Fig.2 of the main text we plot the radial distribution function
along the vertical direction $g_\parallel (r)$ and along the plane $g_\perp(r)$ properly
averaged as follows.
To compute $g_\parallel (r)$ we first divide the plane $x-y$ into small squares and then
compute the radial distribution function following the standard definition 
(see \cite{chaikin1995}) properly normalized along a parallelepiped 
with height given by half the length of the box edge. Finally we average the resulting function
over all the squares in the plane and all the slices of each configuration. 
We repeat this process for $100$ configurations to obtain the curves of Fig.2. 
We proceed analogously for $g_\perp(r)$ where the radial distribution function
is computed along along a parallelepiped with the horizontal basis coinciding 
with the basis of the simulation box.


\begin{thebibliography}{99}
\bibitem{nota4} An extensive calculation of the critical temperature with
of dipolar interactions is not reported here. Indeed it can
be done along the lines of \cite{pilati2008} and it is the focus of a
separated investigation.
\end{thebibliography}
\end{document}